\documentclass[12pt]{iopart}
\usepackage {setstack}
\usepackage{iopams}
\usepackage{amssymb}
\usepackage{amsfonts}
\usepackage {amscd}
\usepackage {latexsym}

\newcommand {\oks}[2]{{\raise0.7ex\hbox{${\scriptstyle #1}$}\!\mathord{\left/
{\vphantom{{1}{2}}}\right.\kern-\nulldelimiterspace}\!\lower0.7ex
\hbox{${\scriptstyle #2}$}}}
\newcommand {\bb}[1]{\mbox{\boldmath $#1$}}

\begin{document}
\title[Neutrino spin rotation in dense matter and electromagnetic field]
{Neutrino spin rotation in dense matter and electromagnetic field}

\author{E. V.
Arbuzova,$^1$ A. E. Lobanov,$^2$ and E. M. Murchikova$^3$}

\address{$^1$ International University ``Dubna'', 141980 Dubna, Russia}
\ead{arbuzova@uni-dubna.ru}
\address{$^2$ Department of Theoretical
Physics, Moscow State University, 119991 Moscow, Russia}
\ead{lobanov@phys.msu.ru}
\address{$^3$ Department of Theoretical
Physics, Moscow State University, 119991 Moscow, Russia}
\ead{murchikova@nanolab.phys.msu.ru}

\pacs{03. 50. De, 12. 20. Ds}


\begin{abstract}
Exact solutions of the Dirac--Pauli equation for massive neutrino
with anomalous magnetic moment interacting with dense matter and
strong electromagnetic field are found. The complete system of
neutrino wavefunctions, which show spin rotation properties are
obtained and their possible applications are discussed.

\end{abstract}

%\maketitle

\section{Introduction}

The neutrino oscillation phenomenon is well established now
\cite{osc.exp.}. For the detailed information about neutrino
flavour oscillations we refer to the pioneer works
\cite{Pontecorvo, MNS}, some resent reviews \cite{osc02, MP} and
references therein. From the theory of this phenomenon it follows
that oscillations are possible only when neutrino possesses
non-vanishing mass. Consequently, a neutrino has non-trivial
electromagnetic properties. In particular, a Dirac neutrino has to
possess a non-zero magnetic moment \cite{LeeShr77}. In turn, the
existence of the magnetic moment is associated with so-called
neutrino spin oscillations. These oscillations are described as
neutrino spin precession or, in other words, as transitions
between left-handed (active) and right-handed (sterile) neutrino
states in an electromagnetic field. Phenomenology of this effect
was widely discussed in the literature \cite{Fu, SchV, VVO, LM,
Akh, BTZ, Smirnov, APS, NSSV97, DM}.

However in the most popular papers the theory of neutrino spin
oscillation was based on solving the Cauchy problem for the
Schr\"{o}dinger type equation with an effective Hamiltonian. So
the states of neutrino were treated as mixed states. The purpose
of our work is to find a complete system of solutions of the wave
equation which accounts for the influence of an electromagnetic
field and a dense matter on neutrino dynamics. The existence of
such system gives a possibility to describe the states of neutrino
with rotating spin as pure states and to evaluate the
probabilities of various processes with neutrino in the framework
of the Furry picture.

\section{Model}

When the interaction of a massive neutrino possessing an anomalous
magnetic moment $\mu_{0}$ with background fermions is considered
to be coherent, its propagation in  matter and electromagnetic
field is described by the Dirac--Pauli equation with the effective
potential \cite{Wolf,MS}. For a moving and polarized matter such
an interaction is described by the effective four-potential
$f^\mu$.

The function
\begin{equation}\label{2}
f^\mu =
{\sum\limits_f{\rho_f^{(1)}j_f^\mu+\rho_f^{(2)}\lambda_f^\mu}}
\end{equation}
is a linear combination of the currents
\begin{equation}\label{3}
j_{f}^\mu=\{n_f u^{0}_f,n_f{\bf{u}}_f\},
\end{equation}
and the polarizations
\begin{equation}\label{4}\lambda^{\mu}_f =\left\{n_f
({\bb{\zeta}}_f{\bf{u}}_f), n_f\left({\bb{\zeta}}_f + \frac{
{\bf{u}}_f ({\bb{\zeta}}_f{\bf{u}}_f)}{1+u^{0}}\right)\right\}
\end{equation}
of background fermions. In these equations $n_f$ and
${\bb{\zeta}}_f \; (0\leqslant |{\bb{\zeta}}_f |^2 \leqslant 1)$
are, respectively, the number density and the mean value of the
polarization vector of the background fermions $f$ in their
center-of-mass system. In this reference frame the mean momentum
of the fermions $f$ is equal to zero, and $u^{\mu}_f
=\{u^{0}_{f},{\bf{u}}_{f}\}$ denote the four-velocity of the
reference frame. Summation is carried out over the background
fermions $f$. We use the units $\hbar =c=1 $.

The expressions for the coefficients $\rho_{f}^{(1,2)}$ depend on
the model chosen for neutrino interactions. In the framework of the
standard model these coefficients calculated in the first order of
the perturbation theory have the forms \cite{101,102,103,104}

\begin{equation}
\rho_f^{(1)}\!=\sqrt 2 {G}_{{\mathrm F}} \left\{
I_{e\nu}+T_3^{(f)}-2Q^{(f)}\sin^2\theta_{\mathrm W}
\right\}\!,\;\; \rho_f^{(2)}\!=\sqrt 2 {G}_{{\mathrm F}} \left\{
I_{e\nu}+T_3^{(f)} \right\}\!.
\end{equation}

\noindent Here $Q^{(f)}$ is an electric charge of the fermion,
$T_{3}^{(f)}$ is a third component of weak isospin, ${G}_{{\mathrm
F}}$ and $\theta_{\mathrm W}$ are the Fermi constant and the
Weinberg angle respectively. For electron neutrino interaction
with electrons \mbox{$I_{e\nu}=1$}, positrons
\mbox{$I_{e\nu}=-1$}, otherwise $I_{e\nu}= 0$.

Let us discuss an equation for neutrino (its mass eigenstate) in
matter and electromagnetic field. In order to avoid correlations
between flavour and spin oscillations (the
Mikheyev--Smirnov--Wolfenstein effect \cite{Wolf,MS}) we shall
suppose that the effective four-potential is the same for all
neutrino flavours and consider the matter, in which a fraction of
electrons is small, i.e. approximately $n_{e}= 0$. In what
follows, we restrict our consideration to the case of a
homogeneous medium. Then the explicit form of the generalized
Dirac--Pauli equation is uniquely determined by the assumptions
similar to those adopted in \cite{F}:
\begin{equation}\label{1}
\left(i\hat {\partial }  - \frac{1}{2}\hat {f}(1 + \gamma ^5) -
\frac{i}{2}\mu_{0}F^{\mu \nu }\sigma _{\mu \nu } - m\right)\Psi =
0.
\end{equation}
\noindent Here $F^{\mu \nu }$ is an electromagnetic field tensor,
$\sigma _{\mu \nu }
=\frac{1}{2}(\gamma^{\mu}\gamma^{\nu}-\gamma^{\nu}\gamma^{\mu})$.
The quantities with hats denote scalar products of the Dirac
matrices with four-vectors: $\hat{a} \equiv \gamma^{\mu}a_{\mu}.$

Note, when propagation of neutrino through a dense matter and an
electromagnetic field is studied, we should keep in mind that the
matter and the field are mainly situated in the same area.
Therefore strengthes of electric and magnetic fields and average
velocities and polarizations of matter should satisfy the
self-consistent system of equations including the Maxwell
equations, the Lorentz equation
\begin{equation}\label{01}
    \dot{u}^\mu_{f}=\frac{e_{f}}{m_{f}}F^{\mu}_{\nu}u^\nu_{f},
\end{equation}
\noindent and the classical spin evolution
Bargmann--Michel--Telegdi (BMT) equation \cite{BMT}
\begin{equation}\label{02}
    \dot{\lambda}^\mu=\left[\frac{e_{f}}{m_{f}}F^{\mu}_{\nu} +2\mu_{f}
    \left( F^{\mu}_{\nu}-u^\mu_{f} F_{\alpha\nu}u^\alpha_{f} \right)
    \right]\lambda^\nu_{f}.
\end{equation}
\noindent Here a dot denotes the differentiation with respect to
the proper time $\tau$.

As already mentioned, we restrict our consideration to constant
velocity and polarization of matter and constant homogeneous
electromagnetic field in equation (\ref{1}). However, even such a
choice imposes certain limitations on $F^{\mu \nu }$ and
$f^{\mu}$. Indeed, function $f^\mu$ is a linear combination of
currents and polarizations of background fermions. Since the
velocities, polarizations and number densities of different
components of matter are independent characteristics in the
general case, then from the condition $f^\mu={\mathrm{const}}$ or
$\dot{f}_{\mu}=0$ it follows that velocities and polarizations of
background components should be stationary: $\dot{u}^\mu_{f}=0,\;
\dot{\lambda}^\mu _{f}=0.$ Using equations (\ref{01}), (\ref{02})
we can obtain the restriction
\begin{equation}\label{03}
F^{\mu\nu}f_\nu=0.
\end{equation}
\noindent It should be stressed that this condition is a direct
consequence of the fact that the average velocity and polarization
of particles of matter in an external field should satisfy the
classical equations of motion.

\section{Calculations}

Let us look for a solution of equation (\ref {1}). Since the
functions $F^{\mu\nu}$, $f^{\mu}$ are constant, the canonical
momentum operator $i{{\partial}_{\mu}}$ commutes with the
Hamiltonian of this equation. However, the commonly adopted choice
of eigenvalues of this operator as quantum numbers is not
satisfactory in our case. Particle kinetic momentum components
related to its group four-velocity $u^{\mu}$ by the relation
$q^{\mu} = mu^{\mu},\; q^{2}= m^{2}$ are more suitable to play
this role. In the present paper we discuss the solutions described
by the quantum numbers which can be interpreted as kinetic
momentum components. The explicit form of a kinetic momentum
operator for a particle with spin is not known beforehand, hence,
in order to find the appropriate solutions, we have to use the
correspondence principle.

It was shown in \cite{L43} that if effects of  neutrino weak
interactions are taken into account, the Lorentz invariant
generalization of the BMT equation for a spin vector $S^{\mu}$ of a
neutrino moving with the four-velocity $u^{\mu}$ has the form
\begin{equation}\label{x4}
{\dot{S}^{\mu}} =2
 \Big\{ \left(F^{\mu\nu}+G^{\mu\nu}\right)S_{\nu} -u^{\mu}
u_{\nu} \left(F^{\nu\lambda}+G^{\nu\lambda}\right)S_{\lambda}
\Big\} ,
\end{equation}
\noindent where
\begin{equation}\label{5}
G^{\mu \nu}= \frac{1}{2}e^{\mu \nu \rho \lambda}
f_{\rho}u_{\lambda}.
\end{equation}
\noindent Here and further for the simplification of the formulae
we include a value of neutrino magnetic moment into an
electromagnetic field tensor, i.e. $\mu_{0}F^{\mu\nu}\Rightarrow
F^{\mu\nu}$.

We introduce quasi-classical spin wavefunctions, which can be
constructed as discussed in \cite{LP,L}. Suppose the Lorentz
equation is solved, and the dependence of particle coordinates on
the proper time is found. Then the BMT equation transforms to an
ordinary differential equation, whose resolvent determines a
one-parametric subgroup of the Lorentz group. The quasi-classical
spin wavefunction is represented by a spin-tensor, whose evolution
is determined by the same one-parametric subgroup. In our case
this spin-tensor represents the Dirac bispinor.

Following the technique presented in the works \cite{LP,SpinLight}
we choose the solution of equation (\ref {1}) in the form

\begin{equation}\label{a2}
\Psi (x) = e^{-iF(x)}U(\tau (x),\tau_{0}=0)\Psi_{0} (x).
\end{equation}

\noindent In this formula $U(\tau,\tau_{0})$ is an operator of
evolution for a quasi-classical spin wavefunction, $e^{-iF(x)}$ is
a phase factor, and

\begin{equation}\label{a3}
\Psi _0 (x) = e^{ - iqx}(1 - \zeta_{0}\gamma ^5\hat {S}_0 )(\hat
{q} + m)\psi _0
\end{equation}

\noindent is a solution of the Dirac equation for a free particle.
Here four-vector $ {S}_0^{\mu}$  determines the direction of
particle polarization, $\zeta_{0} = \pm 1 $ is a sign of spin
projection on this direction and $\psi _{0}$ is a constant bispinor
normalized by the condition $\bar{\Psi}_{0}(x)\Psi _{0}(x)
=m/q_{0}$.

In our case operator $U(\tau,\tau_{0})$ obeys the equation
\begin{equation}\label{a4}
\dot {U}(\tau,\tau_{0}) = \left\{ \frac{i}{4m}\gamma ^5(\hat
{f}\hat {q} - \hat {q}\hat {f}) + \frac{i}{m^{2}}{\gamma ^{5}
H^{\mu \nu }q_{\nu} \gamma _{\mu} \hat {q}}
\right\}{U}(\tau,\tau_{0}),
\end{equation}
\noindent where $H^{\mu \nu } = -
\frac{1}{2}e^{\mu\nu\rho\lambda}F_{\rho\lambda}$ is a  dual
electromagnetic field tensor. It is obvious that the solution of
equation (\ref{a4}) can be represented as a matrix exponent
\begin{equation}\label{a6}
U(\tau,\tau_{0}) = \exp \left\{ {i(\tau-\tau_{0})
\left[{\frac{1}{4m}\gamma ^5(\hat {f}\hat {q} - \hat {q}\hat {f})
+ \frac{1}{m^{2}}\gamma ^5H^{\mu \nu }q_\nu \gamma _\mu \hat {q}}
\right]} \right\}.
\end{equation}

Substitution of the expression (\ref{a2}) into equation (\ref{1})
taking into account (\ref{a4}) leads to the relation
\begin{equation*}\label{aa6}
\begin{array}{l}
\displaystyle\left\{\hat{q} +\hat{\partial}F(x) - \frac{1}{2}\hat
{f} + \frac{1}{2}\gamma^{5} \hat{f} + \gamma^{5} \hat{N} \left[
{\frac{1}{4m}}(\hat{f} \hat{q} - \hat{q} \hat{f}) +
\frac{1}{m^{2}}H^{\mu \nu}q_{\nu} \gamma_{\mu} \hat{q}
\right]\right. \\- \displaystyle\left.\frac{i}{2}F^{\mu \nu
}\sigma _{\mu \nu }-m\right\}e^{-iF(x)}U(\tau(x),\tau_{0})\Psi_{0}
(x)
 = 0,
\end{array}
\end{equation*}
\noindent where $N^{\mu} = \partial ^{\mu} \tau $. Since the
commutator $[\hat{q},U]=0,$ and the matrix $U(\tau(x),\tau_{0})$ is
nondegenerate, then to hold this relation the following condition is
required:
\begin{equation*}\label{a7}
\begin{array}{l}
\displaystyle\hat{\partial}F(x) - \frac{1}{2}\hat {f} +
\frac{1}{2}\gamma^{5} \hat{f} + \gamma^{5} \hat{N} \left[
{\frac{1}{4m}}(\hat{f} \hat{q} - \hat{q} \hat{f}) +
\frac{1}{m^{2}}H^{\mu \nu}q_{\nu} \gamma_{\mu} \hat{q} \right]
\\- \displaystyle\frac{i}{2}F^{\mu \nu }\sigma _{\mu \nu }
 = 0.
\end{array}
\end{equation*}
\noindent Since the coefficients at linearly independent elements
of the Dirac matrix algebra must be equal to zero independently,
we receive
\begin{equation}\label{a8}
    {\partial}^{\mu}F(x) = f^{\mu}/2,
\end{equation}
\noindent and the system of equations for  determination of the
vector $N^{\mu}$:
\begin{equation}\label{106}
\varphi ^\mu (m - (Nq)) + (N\varphi )q^\mu = 0, \quad
     F^{\mu\alpha}q_{\alpha} = - e^{\mu \nu \rho \lambda }
    N_\nu \varphi _\rho q_\lambda,
\end{equation}
\noindent where  $\varphi ^\mu = f^\mu/2 + H^{\mu\nu}q_{\nu}/m$.

It is easily verified that for the compatibility of this system it
is necessary
\begin{equation}\label{dop1}
q_{\mu}F^{\mu\nu}H_{\nu\alpha}q^{\alpha}+mq_{\mu}F^{\mu\nu}f_{\nu}/2=0.
\end{equation}

\noindent Since $q^{\mu}$ takes in general an arbitrary value, in
order to satisfy (\ref{dop1}) identically it is necessary to
demand the conditions
\begin{equation}\label{dop11}
  F^{\mu\alpha}H_{\alpha\nu}\equiv -\frac{1}{4}\delta^{\mu}_{\nu}
  F^{\alpha\beta}H_{\alpha\beta}=0,
\end{equation}
\begin{equation}\label{dop12}
F^{\mu\nu}f_{\nu}=0.
\end{equation}
\noindent Hence, the solution of equation (\ref{1}) can take the
form (\ref{a2}) only  if the tensor $F^{\mu\nu}$ is flat,  i.e.
its second invariant $I_{2}=\frac{1}{4}F^{\mu\nu}H_{\mu\nu}$ is
equal to zero, and the vector $f^{\mu}$ is its eigenvector
corresponding to the zero eigenvalue. It should be noted that the
antisymmetric tensor has an eigenvector corresponding to zero
eigenvalue if and only if this tensor is flat. So conditions
(\ref{dop11}) and (\ref{dop12}) are not independent, and
(\ref{dop11}) follows directly from (\ref{dop12}). Thus we come to
the  condition above (\ref{03}), which was obtained only on the
physical attends.

Let us introduce an orthogonal basis in the Minkowski space

\begin{equation*}\label{Basis}
    %\fl
    \displaystyle n_0^\mu=q^\mu/m, \;\;
    \displaystyle n_1^\mu=\frac{H^{\mu\nu}q_\nu}{\sqrt{{\cal N}}},
    \;\;
    \displaystyle n_2^\mu=\frac{F^{\mu\nu}q_\nu}{\sqrt{\tilde{{\cal N}}}},
    \;\;
    \displaystyle n_3^\mu=\frac{m^{2}H^{\mu\nu}H_{\nu\alpha}
    q^\alpha  -q^\mu {\cal N}}
    {m\sqrt{{\cal N}\tilde{{\cal N}}}},
\end{equation*}

\noindent where ${\cal N}=q_\mu H^{\mu\nu}H_{\nu\rho}q^\rho$,
$\tilde{\cal N}=q_\mu F^{\mu\nu}F_{\nu\rho}q^\rho$. Expanding the
four-vector $N^\mu$ with the basis and using equations (\ref{106})
we get as the result of calculations
\begin{equation}\label{dop111}
\fl\displaystyle N^\mu = n_{0}^{\mu}+ n_{1}^{\mu}\frac{(\varphi
q)\big[{\cal N} -\tilde{\cal N}-m
f_{\mu}H^{\mu\nu}q_{\nu}\big]}{2\sqrt{{\cal N}}((\varphi
q)^2-m^{2}\varphi^2)} + n_{3}^{\mu}\frac{\sqrt{{\tilde{\cal
N}}}}{\sqrt{{\cal N}}}\left[1+\frac{m^{2} (f\varphi) }{2((\varphi
q)^2-m^{2}\varphi^2)}\right].
\end{equation}
\noindent With the help of the relations
\begin{equation}\label{152}
\begin{array}{l}
    2{(\varphi q)}H^{\mu\nu}=m({\varphi^\mu
     f^\nu-f^\mu \varphi^\nu}),\quad {\cal N}-\tilde{\cal N}=2m^{2}
     I_1,\\[8pt]
    ((f\varphi)^2-f^2 \varphi^2)m^{2}=(f^2 {\cal N}
    +(f_{\mu}H^{\mu\nu}q_{\nu})^2)=8(\varphi q)^2 I_{1},
\end{array}
\end{equation}
\noindent where $I_1=\frac{1}{4}F^{\mu\nu} F_{\mu\nu}$ is the
first invariant of the tensor $F^{\mu\nu},$  expression
(\ref{dop111}) can be represented as
\begin{equation}\label{N}
\displaystyle N^\mu = - q^\mu\frac{m(f\varphi)}
    {{2((\varphi q)^2-m^{2}\varphi^2)}}
     + f^\mu\frac{m}{2(\varphi q)}  + \varphi^\mu\frac{m^3(f \varphi)}
    {2(\varphi q)((\varphi q)^2-m^{2}\varphi^2)}.
\end{equation}

Let us note that, in order to  derive formulae of the form
(\ref{152}), one should take the following arguments into
consideration. For an arbitrary antisymmetric tensor $A^{\mu\nu}$,
its dual tensor ${}^{\star\!\!}A^{\mu\nu}=- \frac{1}{2}e^{\mu \nu
\rho \lambda } A_{\rho \lambda}$, and any four-vectors
$g^{\mu},h^{\mu},\;(gh)\neq 0$  the following relation takes place
\begin{equation}\label{A2.16}
A^{\mu\nu}(gh)=
-\left[g^{\mu}A^{\nu\rho}h_{\rho}-A^{\mu\rho}h_{\rho}g^{\nu}\right]
+{}^{\star}\left[h^{\mu}
{}^{\star\!\!}A^{\nu\rho}g_{\rho}-{}^{\star\!\!}
A^{\mu\rho}g_{\rho}h^{\nu}\right].
\end{equation}
\noindent It leads to
\begin{equation}\label{A2.0016}
\begin{array}{l}
g_{\mu}{}^{\star\!\!}A^{\mu}_{\rho}{}^{\star\!\!}
A^{\rho}_{\nu}g^{\nu}h^{2}+\left(g_{\mu}{}^{\star\!\!}A^{\mu}_{\nu}h^{\nu}\right)^{2}
-h_{\mu}A^{\mu}_{\rho}
A^{\rho}_{\nu}h^{\nu}g^{2}-\left(g_{\mu}A^{\mu}_{\nu}h^{\nu}\right)^{2}=
\\[8pt]=g_{\mu}{}^{\star\!\!}A^{\mu}_{\rho}{}^{\star\!\!}
A^{\rho}_{\nu}h^{\nu}(gh)-g_{\mu}A^{\mu}_{\rho}
A^{\rho}_{\nu}h^{\nu}(gh)=2(gh)^{2}I_{1}.
\end{array}
\end{equation}
\noindent Formulae (\ref{152}) follow from equation
(\ref{A2.0016}) if conditions (\ref{dop11}), (\ref{dop12}) are
taken into account.

Since $f^\mu , N^\mu={\mathrm{const}}$ we obtain for the proper time
\begin{equation}\label{xx} \tau = (Nx),
\end{equation}
\noindent and for the phase factor which determines energy shift
of neutrino in matter we have
\begin{equation}\label{xx0}
F(x) =(fx)/2.
\end{equation}
It is not difficult to verify that in view of the relations
obtained, the expression for the wavefunction takes the form
\begin{equation}\label{Wf1}
\displaystyle \Psi(x) = \frac{1}{2}\sum\limits_{\zeta = \pm 1}
e^{-i(P_{\zeta} x)}{( 1 - \zeta \gamma^5
\hat{S}_{tp})(1-\zeta_{0}\gamma^5 \hat{S_0})(\hat{q}+m)} \psi_{0},
\end{equation}
where
\begin{equation}\label{P3}
\displaystyle {S}_{tp}^\mu = \frac{q^\mu (\varphi q) / m -
    \varphi ^\mu m }{\sqrt {(\varphi q)^2 -
    \varphi ^2m ^2}}, \\[12pt]
\end{equation}
\begin{equation}\label{P2}
\begin{array}{l}
\fl\displaystyle P^\mu_{\zeta} = q^\mu + f^{\mu}/2-\zeta
N^{\mu}\sqrt
    {(\varphi q)^2 - \varphi^2 m^2}/m
\displaystyle = q^\mu \bigg(1+\zeta\frac{(f\varphi)}
    {2\sqrt{(\varphi q)^2-m^2\varphi^2}}
    \bigg)\\[12pt]\displaystyle+ \frac{1}{2}f^{\mu}\bigg(1  - \frac{\zeta
\sqrt{(\varphi q)^2-m^2\varphi^2}}{(\varphi q)}\bigg)\,
    \displaystyle -\,\varphi^\mu \,\frac{\zeta (f \varphi) m^2}
    {2(\varphi q)\sqrt{(\varphi q)^2-m^2\varphi^2}}.
\end{array}
\end{equation}

It is obvious that the system of solutions (\ref{Wf1}) is a
complete system of solutions of equation (\ref{1}), which is
characterized by the kinetic momentum of the particle $q^{\mu}$
and the quantum number $\zeta_{0}= \pm 1$, which can be
interpreted as the neutrino spin projection on the direction
${S}_{0}^\mu$ at the moment $\tau = (Nx)=0$. In the general case
this system is not stationary. The received solutions are
stationary only when $S_{0}^{\mu}= S_{tp}^{\mu}$. In this case the
wavefunctions are the eigenfunctions of the spin projection
operator to the direction $S_{tp}^{\mu}$ with the eigenvalues
$\zeta = \pm 1$ and of the canonical momentum operator
$i\partial^{\mu}$ with the eigenvalues $P^{\mu}_{\zeta}$. The
orthonormalized system of the stationary solutions of equation
(\ref{1}) can be written down in the following way:
\begin{equation}\label{Wf}
    \Psi_{\zeta}(x) = e^{ - i(P_{\zeta}x)}\sqrt{|J|}
    (1 - \zeta\gamma ^5\hat {S}_{tp})(\hat {q} + m)\psi_0,
\end{equation}
where $J$ is the transition Jacobian between the  variables
$q^\mu$ and $P^\mu_{\zeta}$:
\begin{equation*}\label{nor3}
  J=  {\mathrm{det}}(M_{ij}) =
    {\mathrm{det}} \left[ \frac{\partial P^i_{\zeta}}{\partial
    q^j}+\frac{\partial P^i_{\zeta}}{\partial q^0} \frac{\partial q^0}{\partial
    q^j} \right].
\end{equation*}
\noindent Using formulae (\ref{152}) it is possible to rewrite the
matrix $M_{ij}$ in the form
\begin{equation*}
\begin{array}{l}
    \displaystyle M_{ij}=\delta_{ij}
    \left( 1+\zeta\frac{(f\varphi)}{2\sqrt{(\varphi q)^2-m^2 \varphi^2}}
    \right)+\\[20pt]
    \displaystyle+\zeta\left( q_i-\varphi_i\frac{m^2}{(\varphi q)} \right)
    \left( \varphi_j-q_j\frac{\varphi^0}{q^0} \right)\frac{m^2(f \varphi)^2+f^2
    (\varphi q)^2-m^2 f^2\varphi^2}{ 4(\varphi q)((\varphi
    q)^2-m^2\varphi^2)^{3/2}}.
\end{array}
\end{equation*}
\noindent It can be easily shown that for any vectors ${\bf g}$
and ${\bf h}$ the relation
$$\det{(\delta_{ij}+g_{i}h_{j})}=1+({\bf{gh}})$$ is correct. Hence
\begin{equation}\label{Det}
    J={\left({1+\zeta\frac{(f \varphi)}
    {2\sqrt{(\varphi q)^2-m^2 \varphi^2}}}\right)}^2
    {\left({1+\zeta\frac{f_{\mu}H^{\mu\nu}q_{\nu}/2m-2
    I_1}
    {\sqrt{(\varphi q)^2-m^2 \varphi^2}}}\right)}.
\end{equation}

The structure of  solution (\ref {Wf}) directly leads us to the
conclusion that when neutrinos move through a dense matter and an
electromagnetic field which satisfy condition (\ref{dop12}), they
can behave as free particles,  i.e. move with the constant group
velocity
\begin{equation}\label{gr}
{\bf v}_{gr} = \frac{\partial {P}_{\zeta}^{0}}{\partial {\bf
    P}_{\zeta}}= \frac{{\bf
    q}}{{q}^{0}}
\end{equation}
\noindent and conserve the polarization.

However in interactions with other particles the channels of
reactions which are closed for a free neutrino can be opened (see,
for example, \cite{SpinLight, zlm}), as a result of difference of
the dispersion law for the free neutrino $P^2=m^2$ and the one for
neutrino in matter and electromagnetic field:
\begin{equation}\label{PP3}
    \tilde{P}^2=m^2-f^{2}/4-2 I_1-2\zeta \Delta\sqrt{(\tilde{P}
    \tilde{\Phi})^2-\tilde{\Phi}^{2}m^2},
\end{equation}
where
\begin{equation}\label{bbb}
\begin{array}{c}
 \displaystyle   \tilde{P}^\mu=P^{\mu}_{\zeta}-f^{\mu}/2, \quad
    \tilde{\Phi}^\mu=f^\mu/2+H^{\mu\nu}\tilde{P}_\nu/{m},\\[8pt]
    \displaystyle\Delta =
{\mathrm{sign}}{\left({1+\zeta\frac{f_{\mu}H^{\mu\nu}q_{\nu}/2m-2
    I_1}
    {\sqrt{(\varphi q)^2-m^2 \varphi^2}}}\right)}.
\end{array}
\end{equation}
\noindent In order to get equation (\ref{PP3}) we used the
following relations obtained from formulae (\ref{A2.16}) and
(\ref{A2.0016})
\begin{equation*}\label{vvv}
\begin{array}{c}
\displaystyle  (\tilde{P}f)= 2(\varphi
q)\left({1+\zeta\frac{f_{\mu}H^{\mu\nu}q_{\nu}/2m-2
    I_1}
    {\sqrt{(\varphi q)^2-m^2 \varphi^2}}}\right),\\[14pt]
    \displaystyle (\tilde{P} f)(f\varphi ) = 2(\varphi q)(f\tilde{\Phi}
    ),\quad {\tilde{\Phi}}^{2}(\varphi q)^{2} = \varphi^{2}(\tilde{\Phi}
    \tilde{P})^{2}.
\end{array}
\end{equation*}
\noindent The appearance of the factor $\Delta$ in equation
(\ref{PP3}) is connected with the fact that $\zeta$ is a particle
spin projection on the direction determined by a kinetic particle
momentum instead of the canonical one.

Note that the solutions which are classified by a kinetic momentum
have been earlier found for equation (\ref{1}) in which only
$F^{\mu\nu}\neq0$ or only $f^{\mu}\neq0$ in papers \cite{LP} and
\cite{SpinLight} respectively. These solutions can be received from
formula (\ref{Wf1}) (the normalized stationary solutions from
formula (\ref{Wf})), if either $f^{\mu}= 0$ or $F^{\mu\nu}=0$ .

\section{Discussion}

Let us discuss  the physical meaning of the results obtained. For
this purpose we shall consider vector and axial currents
constructed with help of solution (\ref{Wf1}). The vector current
is
\begin{equation}\label{g1}
V^{\mu}=\bar{\Psi}(x)\gamma^{\mu}{\Psi}(x)=q^{\mu}/q^{0},
\end{equation}
i.e.  solution (\ref{Wf1}) describes neutrino propagating with the
velocity ${\bf{v}} ={\bf{q}}/q^{0}$. At the same time the axial
current is
\begin{equation}\label{ggg1}
A^{\mu}=\bar{\Psi}(x)\gamma^{5}\gamma^{\mu}{\Psi}(x)=\zeta_{0}\frac{m}{q^{0}}S^{\mu}.
\end{equation}
Here
\begin{equation}\label{g2}
S^{\mu}=-S_{tp}^{\mu}(S_{0}S_{tp})+\!\left[S_{0}^{\mu}+S_{tp}^{\mu}(S_{0}S_{tp})\right]
\cos2\theta
-\frac{1}{m}e^{\mu\nu\rho\lambda}q_{\nu}S_{0\rho}S_{tp\lambda}\sin2\theta,
\end{equation}
where
\begin{equation}\label{g3}
\theta = (Nx)\sqrt {(\varphi q)^2 -
    \varphi^2 m^2}/m.
\end{equation}
\noindent The spin vector $\bb\zeta$ can be expressed in  terms of
the four-vector $S^{\mu}$ components as
\begin{equation}\label{ad201}
   \bb{\zeta} ={\bf{S}}-\frac{{\bf{q}}S^0}{q^{0}+m}.
\end{equation}
\noindent As the consequence of this fact we have
\begin{equation}\label{ad201x}
   \bb{\zeta} ={\bb \zeta}_{tp}({\bb \zeta}_{0}{\bb \zeta}_{tp})+
   [{\bb \zeta}_{0}- {\bb \zeta}_{tp}({\bb \zeta}_{0}{\bb \zeta}_{tp})]\cos2\theta
   -[{\bb \zeta}_{tp}\times
   {\bb \zeta}_{0}]\sin2\theta.
\end{equation}

Introduce a flight length $L$ of a particle and an oscillations
length $L_{{osc}}$, using the relation $\theta =\pi L/L_{{osc}}.$
Since the scalar product $(Nx)= \tau$ can be interpreted as the
proper time of a particle, then the oscillation length is defined
as
\begin{equation}\label{g4}
L_{{osc}} = \frac{2\pi{\hbar} c\,{|\bf{q}|}}{\sqrt {(fq)^2 - f^{2}
m^2 c^2 - 4m
c\mu_{0}H^{\mu\nu}b_{\mu}q_{\nu}+4\mu_{0}^{2}H^{\mu\alpha}
H_{\alpha\nu}q_{\mu}q^{\nu}}}.
\end{equation}
\noindent In this formula we use gaussian units and restore the
neutrino magnetic moment $\mu_{0}$.

Hence if as a result of a certain process a neutrino arose with
polarization ${\bb \zeta}_{0}$, after travelling the distance $L$
the probability for the neutrino to have polarization ${-\bb
\zeta}_{0}$ is equal to
\begin{equation}\label{g14}
    W_{sf}= \big[{\bb{\zeta}}_{0}\times{\bb {\zeta}}_{tp}\big]^{2}
    \sin^{2}(\pi L/L_{{osc}}).
\end{equation}
Consequently, if the condition $({\bb{\zeta}}_{0}{\bb
{\zeta}}_{tp})=0$ is fulfilled, this probability can become unity,
i.e. a resonance takes place.

Thus, solution (\ref{Wf1}) which is a linear combination of
solutions (\ref{Wf}) describes a spin-coherent state of neutrino,
in which a spin rotation takes place. Therefore, neutrino state
with rotating spin is a pure state. Existence of such solutions is
the direct consequence of the neutrino state description in terms
of kinetic momentum. It should be stressed that as the result of
calculations we obtained the complete system of neutrino
wavefunctions, which show spin rotation properties.

Such a spin behaviour has a simple quasi-classical interpretation.
The antisymmetric tensor $G^{\mu \nu}$ (see equation (\ref{5}))
can be written in the standard form
\begin{equation}
G^{\mu \nu}= ({\bf P}, {\bf M}\,), \label{g6}
\end{equation}
where
\begin{equation}\label{g7}
{\bf M}= (f^{0} {\bf q}-q^{0}{\bf f})/(2m), \quad {\bf P}=-
\big[{\bf q \times {\bf f}}\big]/(2m).
\end{equation}
\noindent Vectors ${\bf P}$ and ${\bf M}$ are analogous to the
polarization and the magnetization vectors of matter. It should be
noted that the substitution $F^{\mu\nu}\Rightarrow F^{\mu\nu}+G_{\mu
\nu}$ implies that the magnetic ${\bf H}$ and electric ${\bf D}$
fields are shifted by the vectors ${\bf M}$ and ${\bf P}$,
respectively:
\begin{equation}\label{g8}
{\bf H} \Rightarrow {\bf B} ={\bf H} +{\bf M}, \quad {\bf D}
\Rightarrow {\bf E}= {\bf D} -{\bf P}.
\end{equation}
\noindent So restriction (\ref{03}) in the explicit form is
\begin{equation}\label{fields}
({\bf{E}}{\bf{f}})=0, \quad
  {\bf{E}}f^0-[{\bf{B}}\times{\bf{f}}]=0.
\end{equation}

In the rest frame of the particle, equation (\ref{x4}) can be
written down in the form
\begin{equation}\label{g11}
\dot{{\bb \zeta}}={2} \big[ {{\bb\zeta} \times {\bf B}_0} \big],
\end{equation}
\noindent where the spin vector $\bb\zeta$ is related to
four-vector $S^{\mu}$ by equation (\ref{ad201}) and the value
${\bf B}_0$  is the effective magnetic field in the neutrino rest
frame  which can be expressed in terms of quantities determined in
the laboratory frame
\begin{equation}\label{g12}
\begin{array}{l}
{\bf B}_0= \displaystyle\frac{1}{m}\left[q^{0}{\bf B} -\big[{\bf
q}\times{\bf E}\big]-\frac{{\bf q}({\bf q}{\bf
B})}{q^{0}+m}\right]\\[12pt]\displaystyle =\frac{1}{m}\left[q^{0}{\bf H}
-\big[{\bf q}\times{\bf D}\big]-\frac{{\bf q}({\bf q}{\bf
H})}{q^{0}+m}+\frac{\bf q}{2} \Big(f^{0}-\frac{({\bf q}{\bf
f})}{q^{0}+m}\Big)\right] -\frac{\bf f}{2}.
\end{array}
\end{equation}
Thus, the neutrino spin precesses around the direction ${\bf B}_{0}$
with the frequency $\omega = 2m|{\bf B}_{0}|/q^{0} =2\pi|{\bf q}|
/(q^{0}L_{{osc}}).$ It is not difficult to see that the spin vector
direction corresponding to stationary states $\bb{\zeta}_{tp}$ is
connected to the effective magnetic field as follows:
\begin{equation}\label{g13}
{\bb{\zeta}}_{tp}=\frac{{\bf B}_0}{|{\bf B}_0|}=
\frac{mL_{{osc}}}{\pi|{\bf{q}|}}{\bf B}_0.
\end{equation}
\noindent This fact explains the stationarity of states with
$S^{\mu}_{0}=S^{\mu}_{tp}$.

If neutrino possesses the fixed helicity in the initial state,
i.e.
\begin{equation}\label{g003}
S_{0}^{\mu} =  S_{sp}^{\mu}=\frac{1}{m}\left\{{|\bf q|}, q^{0}{\bf
q}/{|\bf q|}\right\},\quad {\bb{\zeta}}_{0}={\bb{\zeta}}_{sp}=
\frac{{\bf q}}{|{\bf q}|},
\end{equation}
\noindent then formula (\ref{g14}) simplifies to the result widely
discussed in the literature, which was mentioned in Introduction.

Clearly, our arguments based on the consideration of vector and
axial currents and the final formula (\ref{g14}) are illustrations
for neutrino propagation through background media. As it is
commonly known \cite{LanP}, in a relativistic case only the
integrals of motion are well defined observables.

\section{Conclusions}

In this way we obtained the exact solutions of the Dirac--Pauli
equation for neutrino in dense matter and electromagnetic field.
It was demonstrated that if the neutrino production occurs in the
presence of an external field and a dense matter, then its spin
orientation is characterized by the vector $S_{tp}^{\mu}$ instead
of the vector $S_{sp}^{\mu}$. Using both the stationary (\ref{Wf})
and the nonstationary (\ref{Wf1}) solutions it is possible to
calculate the probabilities of various processes with neutrino in
the framework of the Furry picture. Due to the time--energy
uncertainty relation the considered states of neutrino can be
generated only when the linear size of the area occupied by the
electromagnetic field and the matter is comparable with the
process formation length. This length is of the order of the
oscillations length.

\ack

The authors are grateful to A. V. Borisov, O. F. Dorofeev and V.
Ch. Zhukovsky  for helpful discussions. This work was supported in
part by the grant of President of Russian Federation for leading
scientific schools (Grant SS --- 5332.2006.2).

\end{document}